\documentclass[a4paper,11pt]{article}
\usepackage{pos}
\usepackage{graphicx}
\usepackage{siunitx}
\usepackage{upgreek}
\DeclareSIUnit\parsec{pc}

\title{Design and performance of the multi-PMT optical module for IceCube Upgrade}

\ShortTitle{mDOM Design and Performance}

\author{The IceCube Collaboration \\{\normalsize \normalfont(a complete list of authors can be found at the end of the proceedings)\\}}
\author{K.-H.\ Sulanke\\[3mm]}

\emailAdd{lew.classen@icecube.wisc.edu}
\emailAdd{tba109@psu.edu}
\emailAdd{atfienberg@psu.edu}
\emailAdd{sarah.mechbal@desy.de}
\emailAdd{judith.schneider@desy.de}
\emailAdd{karl-heinz.sulanke@desy.de}
\emailAdd{m\_unla02@uni-muenster.de}
\emailAdd{chris.wendt@icecube.wisc.edu}

\abstract{
The IceCube Upgrade is the first step towards the next-generation neutrino observatory at the South Pole, IceCube-Gen2, and will be installed in the central region of the existing array. The Upgrade will  consist of 693 newly developed, densely spaced optical sensors and 50 standalone calibration devices, which will enhance IceCube's capabilities both at low and high neutrino energies. Of the new sensors, 402 will be multi-PMT Digital Optical Modules (mDOMs). Consisting of 24 small photomultipliers arranged inside a pressure vessel, the mDOM features a large sensitive area distributed nearly homogeneously over the full solid angle. The use of multiple, individually read-out PMTs allows directional information to be obtained for the registered photons and enables the use of multiplicity triggering within a single module, e.g., for background suppression. The challenges driving the mDOM development included tight restrictions on module size, data-transfer rate, and power consumption as well as the harsh environment in the deep ice at the South Pole. In this contribution we present the final mDOM design that meets these challenges.\\

\vspace{4mm}
{\bfseries Corresponding authors:}
T.\ Anderson$^{1}$, L.\ Classen$^{2*}$, A.T.\ Fienberg$^{3}$, S.\ Mechbal$^{4}$, J. Schneider$^5$, K.-H.\ Sulanke$^4$, M.A.\ Unland~Elorrieta$^{2}$, C.\ Wendt$^{6}$\\

{$^{1}$ \itshape Dept. of Physics, Pennsylvania State University, University Park, USA}\\
{$^{2}$ \itshape Institut f\"ur Kernphysik, Westf\"alische Wilhelms-Universit\"at M\"unster, M\"unster, Germany}\\
{$^{3}$ \itshape Dept. of Physics, Pennsylvania State University, University Park, USA}\\
{$^{4}$ \itshape DESY, Zeuthen, Germany}\\
{$^{5}$ \itshape Erlangen Centre for Astroparticle Physics, Erlangen, Germany }\\
{$^{6}$ \itshape Dept. of Physics and Wisconsin IceCube Particle Astrophysics Center, University of Wisconsin{\textendash}Madison, Madison, USA }\\[4mm]
$^*$ Presenter

\FullConference{37$^{\rm{th}}$ International Cosmic Ray Conference (ICRC 2021)\\
		July 12th -- 23rd, 2021\\
		Online -- Berlin, Germany}
}

\begin{document}
\maketitle

\section{Introduction}
\label{sec:mdom_general}
Located deep in the ancient glacial ice of Antarctica, IceCube \cite{Aartsen2016:icecube} is the neutrino telescope with the largest instrumented volume worldwide. IceCube's initial energy range, optimized for the investigation of the neutrino sky at the $\si{\tera\electronvolt}$ to $\si{\peta\electronvolt}$ energy scale and beyond, was extended down to $\SI{\sim 10}{\giga\electronvolt}$ by DeepCore \cite{Abbasi2012:deepcore}, enabling world-class measurements of neutrino oscillation parameters \cite{PhysRevD.99.032007, PhysRevLett.120.071801}. The IceCube Upgrade \cite{Ishihara2019:IC_Upgrade} will further enhance IceCube's capabilities through the installation of 693 new optical modules distributed along seven vertical strings, mainly located in the DeepCore region. The Upgrade will reduce IceCube's energy threshold to a few $\si{\giga\electronvolt}$ which will significantly enhance the precision of oscillation measurements. It will also provide a platform for improved calibration of the existing detector. The enhanced understanding of the optical properties of the deep ice will reduce the main systematic errors that contribute to the directional uncertainty of astrophysical neutrinos allowing us to re-analyze more than ten years of archival IceCube data.

The optical module is the basic building block of a large-volume neutrino telescope. In first-generation detectors it features a single large PMT measuring the amount of incoming photons (derived from the signal charge) as well as their arrival times. Novel optical sensors will play a key role in the expected performance enhancements of the IceCube Upgrade. A large fraction will be multi-PMT Digital Optical Modules (mDOMs) featuring 24 relatively small PMTs (see Sec.~\ref{sec:pmts}). This multi-PMT approach, introduced to deep-sea detectors by the KM3NeT Collaboration \cite{Loehner2013:km3net_module}, results in attractive advantages with respect to the traditional single-PMT technique, including a homogeneous solid angle coverage and a larger sensitive photocathode area per module. Furthermore, the mDOM provides not only the number and arrival time of photons, but also directional information as well as the possibility of multiplicity triggering inside one module.

\section{mDOM Design: Development and Status}
\begin{figure}[tb]
\begin{minipage}{0.5\textwidth}
    \centering  
   \includegraphics[width=0.6\textwidth]{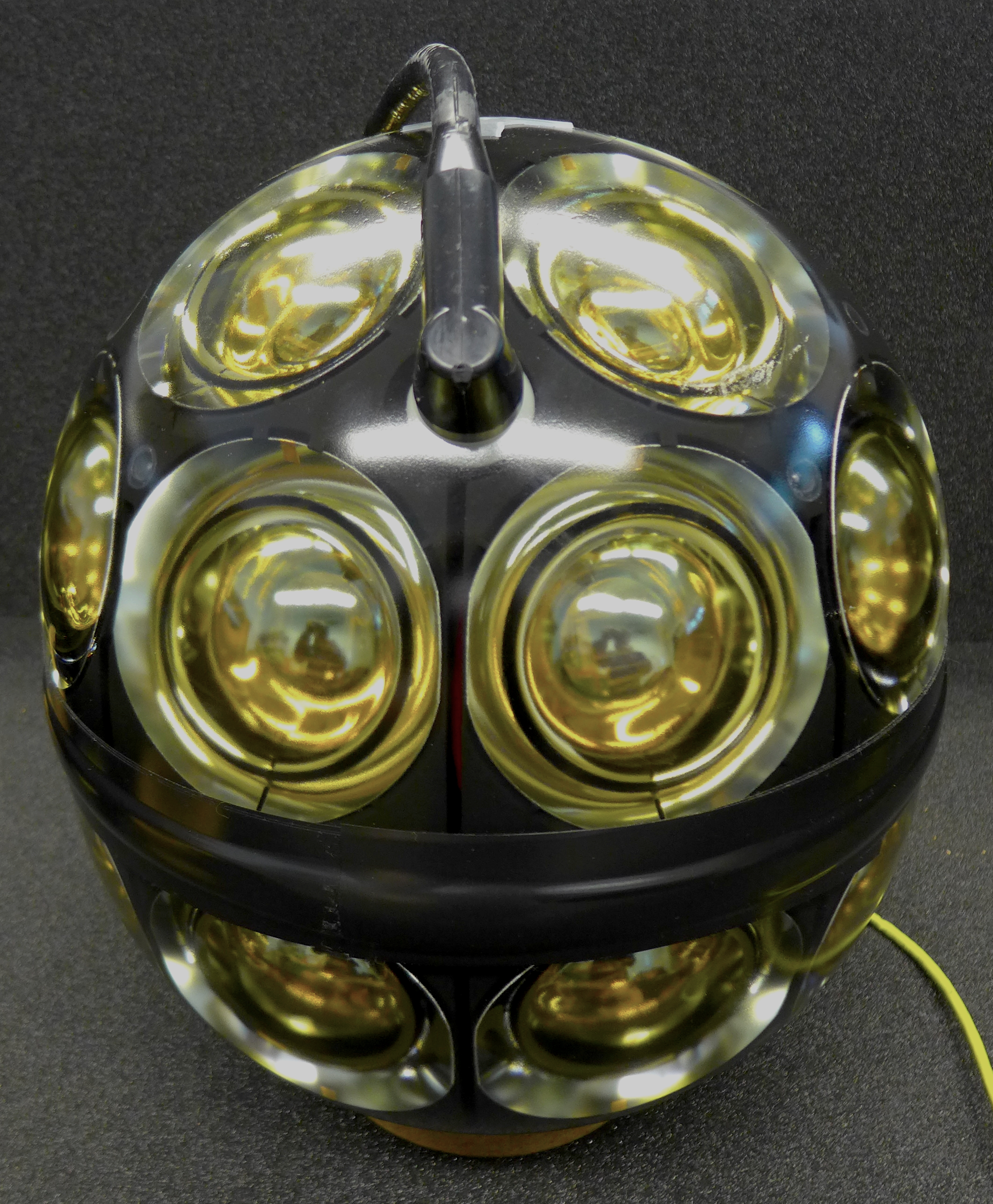}
\end{minipage}
\hfill
\begin{minipage}{0.5\textwidth}
    \centering
    \includegraphics[trim = 14mm 0 68mm 0, clip, width=\textwidth]{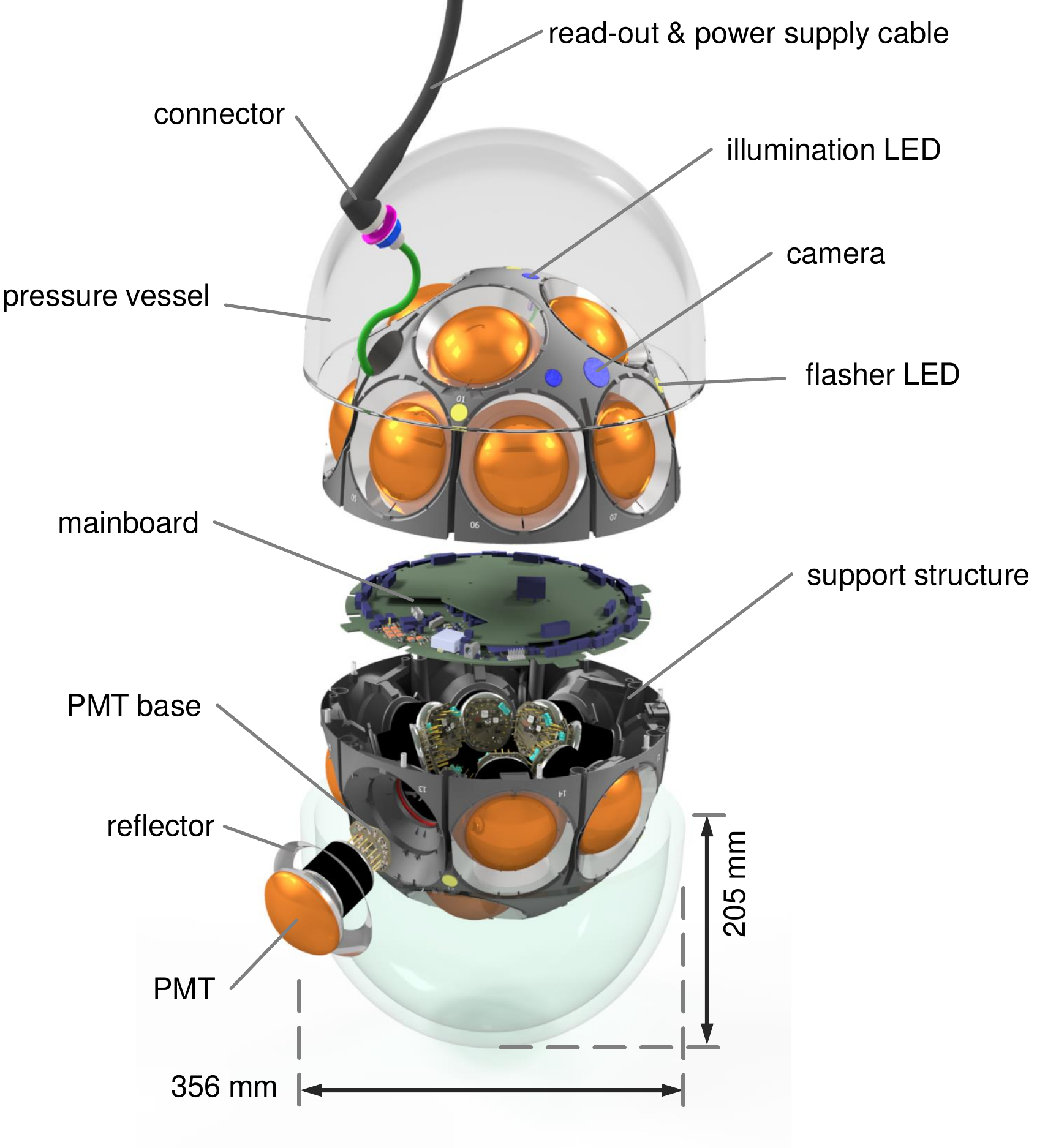}
\end{minipage}
\caption{mDOM overview: \emph{Left}: mDOM constructed for the Design Verification Test (DVT) campaign \emph{Right}: Exploded view featuring main components.}
\label{fig:mDOM}
\end{figure}

The environmental conditions in the deep ice and the detector infrastructure at the South Pole pose unique challenges to optical module technology, such as a limited borehole diameter, pressure spikes during freeze-in and a tight power budget per module. The mDOM design was driven by these challenges. Figure~\ref{fig:mDOM} shows a picture of one of the first fully assembled mDOMs together with an exploded view highlighting different components. The mDOM features 24 three-inch class PMTs\footnote{The actual PMT diameter is $\SI{80}{mm}$ ($\sim3.15$ inch).} each equipped with its own active base that generates the high voltage in-situ. The PMT signals are routed to the mainboard where they are digitized, processed in the mainboard's central FPGA and sent to the surface via the ICM (Ice Communication Module) which is the communication  interface common to all devices in the ice. The PMTs as well as calibration devices are fixed in place with a 3D-printed support structure which is glued and optically coupled with silicone-based gel to the pressure housing made from borosilicate glass. In the following we give a more detailed account of the key components. 

\subsection{Photomultipliers}
\label{sec:pmts}
The PMT used in the mDOM is the R15458-02 model by Hamamatsu. Based on the R12199-01 HA MOD, it was further optimized for the tight spatial constraints inside this module, resulting in a reduced overall tube length of $\SI{91}{mm}$. 
The main characteristics of the R15458-02 model were not affected by the reduced length and are comparable to the R12199-01 HA MOD, which was characterized in detail in  \cite{Unland2019:pmts}. 
The PMTs come equipped with an active base (see Sec.\ref{sec:base}) by the manufacturer, with the PMT-base compound referred to as R15458-20.
The performance of all PMTs used for mDOM production will be tested in a dedicated facility with mass-testing capabilities \cite{icrc2021:pmt_testing}.

\subsection{Active PMT Bases}
\label{sec:base}
\begin{figure}[tb]
    \centering
    \includegraphics[width=1\textwidth]{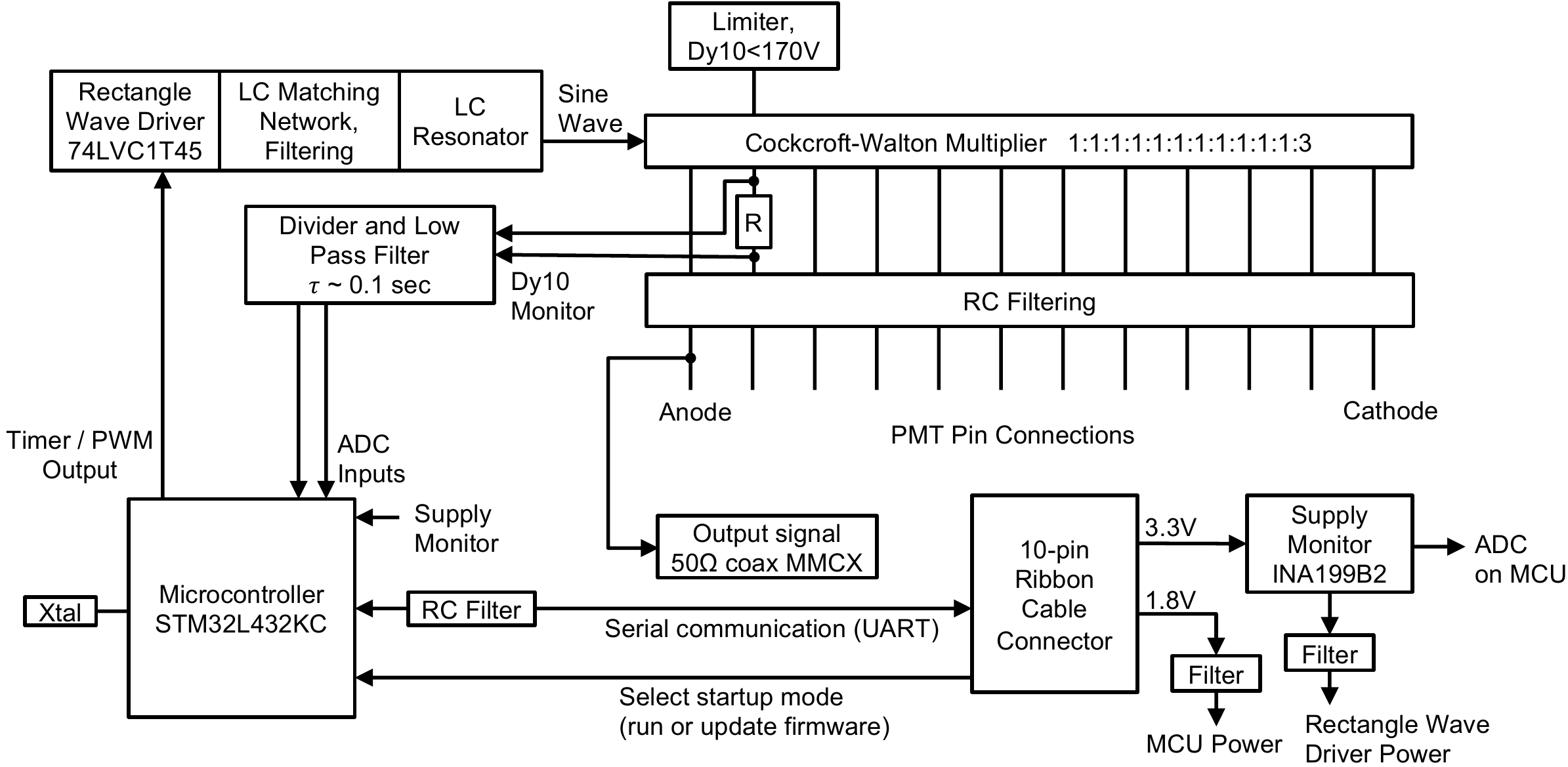}
\caption{Block diagram of the active PMT base.}
\label{fig:uBase}
\end{figure}
The PMT bases were custom designed for the mDOM PMT, featuring low power, off-the-shelf components, and dimensions to fit the very limited space in the center of the mDOM. Figure~\ref{fig:uBase} shows the block diagram.  The external connections are via a ribbon cable with low voltage power and a UART (Universal Asynchronous Receiver Transmitter) command interface, and a coax signal output cable connected directly to the anode.

To generate and regulate high voltage, a microcontroller on each base creates a rectangular waveform with adjustable frequency and duty cycle, switching between 0 and 3.3V at \SI{100}{kHz}--\SI{110}{kHz} (typical).  This waveform drives an LC resonator to output a sine wave with peak-to-peak amplitude \SI{75}{V}--\SI{150}{V}, which drives a 13-stage Cockcroft-Walton multiplier.  In contrast to previously reported PMT bases with resonant drive circuits (e.g., Ref.~\cite{Crisler2010:COUPPBase}), the resonator is direct coupled with no transformer.  The inter-dynode voltages starting from the anode follow approximately the recommended ratios 1:1:1:1:1:1:1:1:1:1:3, ending with the cathode at negative high voltage.  

The voltage is adjusted by changing the duty cycle of the switched rectangular drive waveform, with a fixed frequency below the resonance maximum. 
The control is optimized when the  frequency is chosen so $V_\mathrm{max}$ is 5\%-10\% above the desired $V_\mathrm{out}$ value. The microcontroller program accepts voltage setting commands via the UART interface, automatically determines an appropriate operation frequency and monitors performance.  New versions of the firmware can be uploaded via the UART.

For regulation, a control loop is implemented in the microcontroller program, using the first stage (dynode 10, Dy10) voltage as the measured quantity and the duty cycle as the controlled quantity.  Successive stage voltages are slightly smaller so the cathode is at $\sim$12 times the Dy10 voltage. The loop dynamics include proportional and integral error terms, with measurements taken 10 times per second.  The output time constant is about \SI{1}{s} and stability is typically 0.03\% (RMS).  

The power for the high voltage generator for an output of $1200\,{\rm V}$ is typically $5\,{\rm mW}$.
The power consumption is dominated by losses in the resonance inductor. 
The microcontroller operates from $1.8\,{\rm V}$ and adds another $2\,{\rm mW}$.

The resonator drive circuit includes an impedance matching LC pair to minimize the switch current and corresponding high frequency transients (EMI).
  
RC filters are used between Cockcroft-Walton outputs and the dynodes and cathode, with an additional filtering layer for dynodes near the anode output.  
The final ripple and switching noise are typically below $20\,\mathrm{\upmu Vpp}$ as observed at the anode output (with $50\,\Omega$ load).  

The bases were manufactured under contract with Hamamatsu, and tested at the factory using equipment and procedures supplied by the IceCube collaboration. This testing included cycling between $\SI{-40}{\celsius}$ and $\SI{85}{\celsius}$ and a ``burn-in'' period at $\SI{85}{\celsius}$. It also included subsequent measurement of all dynode output voltages for accuracy and stability.  After soldering to PMTs, Hamamatsu tested the combined assemblies for correct signal output before shipping.

\subsection{mDOM Mainboard}

\begin{figure}[tb]
    \centering
    \includegraphics[width=0.75\textwidth]{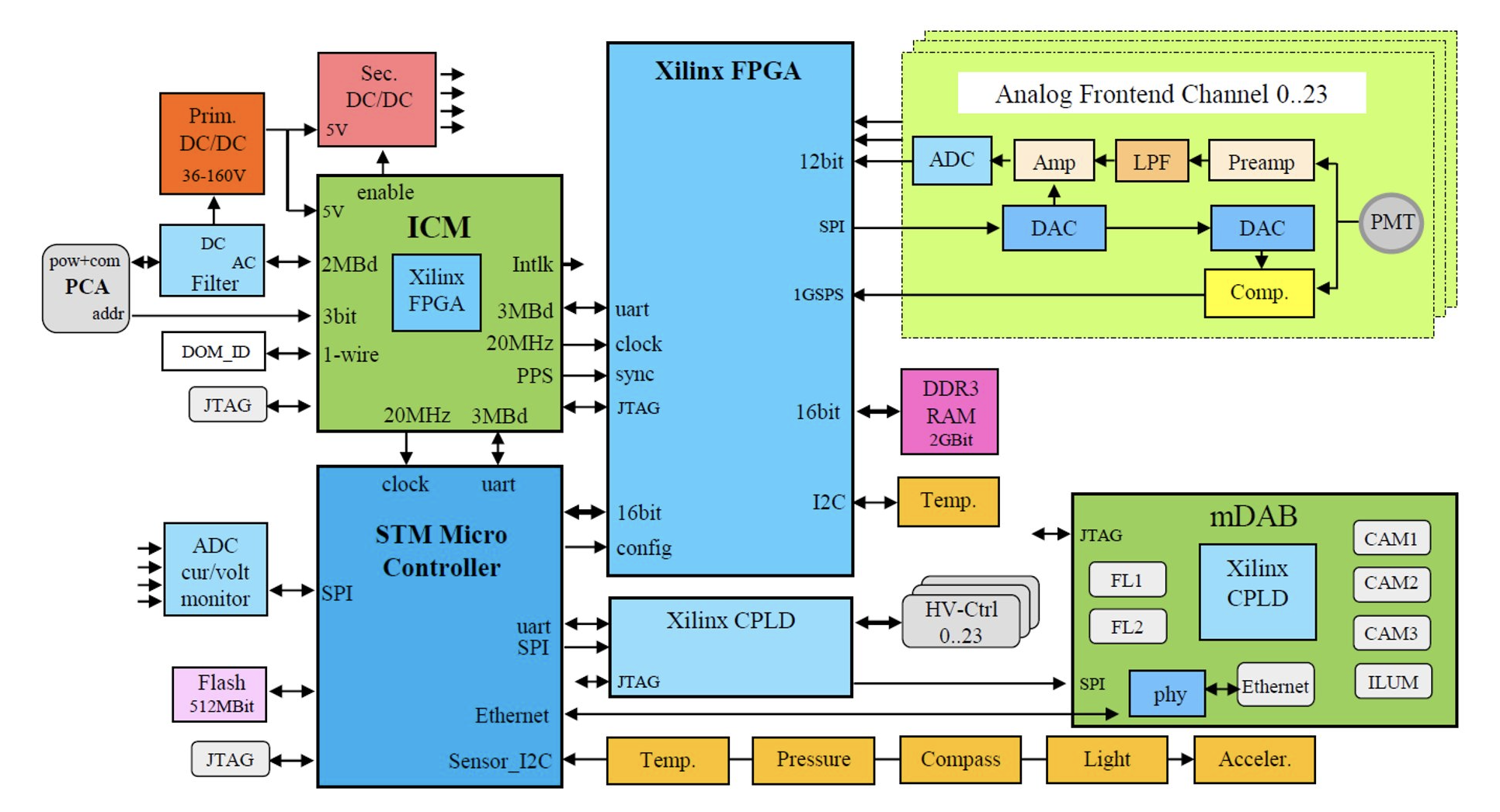}
\caption{Simplified block diagram of the mDOM mainboard.}
\label{fig:mb_block_diagram}
\end{figure}
The main parts of the mDOM mainboard are two DC/DC converters, circuitry to monitor currents and voltages, 24 AFE (analog front-end) channels, a \SI{2}{Gbit} DDR3 RAM based event buffer, a powerful FPGA, a MCU (Microcontroller Unit), the ICM (IceCube Communication Module), a Xilinx CPLD, various sensors and the mDAB (mDOM Adapter Board). A simplified block diagram is depicted in Fig.~\ref{fig:mb_block_diagram}. The overall power consumption measured at the input of the primary DC/DC is $\sim \SI{10}{\watt}$. 

Communication signals and DC power are provided to the mDOM by a single wire pair. A precision \SI{20}{\mega\hertz} oscillator provides a system clock to both the MCU and the FPGA. Data exchange, mainly between the ICM and the MCU, takes place through a UART channel, running at \SI{3}{MBd}. Various sensors, most of them located at the border of the PCB, can be read out using a I2C bus connection. The MCU initializes the AFE channels and sets and monitors the individual operation voltages of the PMTs.  
The MCU's SPI bus is used to individually power-enable any of the AFE channels. The same bus also controls the mDAB board. The mDOM adapter board, equipped with a Xilinx CPLD, is primarily needed to accommodate the connectors of the three attached cameras, their illumination board and the two flasher LED chains. 

The central task of the main FPGA is to control the 24 AFE channels. The ADC baselines and trigger thresholds are set by the MCU through the FPGA. Each of the 24 PMT channels provides its ADC and discriminator outputs to the mainboard FPGA. The \SI{12}{bit} ADCs are operated at \SI{120}{\mega\hertz}, and the discriminators are sampled at \SI{960}{\mega\hertz}. The digitized waveforms along with the associated discriminator signals allow for precise leading-edge time extraction (resolution of about \SI{1}{ns}). Custom firmware inspects each of the 24 datastreams for trigger conditions in real time. Waveform acquisition can be triggered either by the discriminator or by over-threshold ADC samples. Following a trigger, the FPGA writes the trigger sample along with a configurable number of pre- and post-samples into an internal buffer. Triggers in one channel do not interfere with triggering logic in any other channel. The waveform acquisition is deadtime-free provided the FPGA's internal buffers do not overflow.

Data containing ADC waveforms and discriminator samples are transferred out of the waveform buffers and into DDR3 SDRAM by the FPGA. This process can occur at the same time as new waveforms are being written into the buffer. The mainboard has \SI{2}{gigabits} of DDR3 memory, which is enough to buffer many seconds or even minutes of PMT data. This amount of data storage is sufficient to deal with the trigger delay caused by the surface-based trigger processing.

The most challenging aspect of the AFE design was to accomplish low power consumption while maintaining enough bandwidth, sufficient dynamic range and good linearity. In contrast to traditional designs, a fully DC-coupled approach was used in order to achieve low noise in the neighborhood of electromagnetic interference. This avoids any droop effects (ADC-baseline variations) depending on the time between consecutive PMT pulses. The PMT signal, connected to the discriminator through a $\SI{49.9}{\ohm}$ resistor, is terminated at a $\SI{0.95}{\volt}$ DC level. Two precise, \SI{16}{bit} DAC channels are used to adjust the discriminator threshold and the ADC baseline. For pulse shaping, two low pass filters are being used, one between the two amplifier stages and one in front of the ADC. The precision ($0.1\%$) gain setting resistors are chosen to achieve a dynamic range of about $\SI{70}{pe}$ (photoelectron).

In order to characterize the behavior of the bare AFE channels, parameters such as the signal-to-noise-ratio (SNR), dynamic range, linearity as well as long-term stability of the discriminator threshold were investigated at low temperatures using a one-channel piggy-back board with the mainboard AFE design connected to a calibrated PMT, illuminated by tunable pulsed LED, together with a second reference PMT read-out by a digitizer.

\begin{figure}[tb]
\begin{minipage}{0.7\textwidth}
    \centering  
     \includegraphics[trim = 0 0 0 0, clip,width=0.95\textwidth]{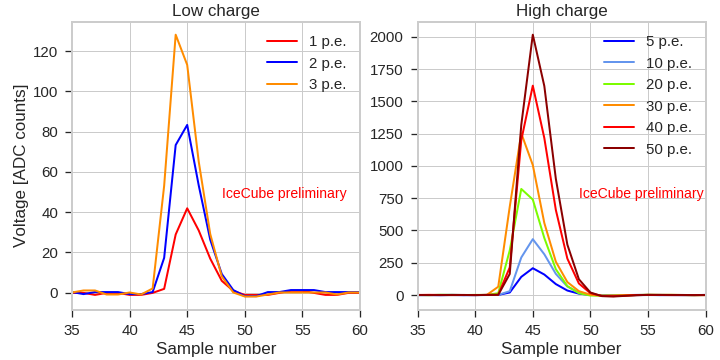}
\end{minipage}
\begin{minipage}{0.25\textwidth}
    \caption{Example PMT pulses (1pe = single photoelectron) acquired at $\SI{-22}{\celsius}$ with a time sampling of 100 MSPS and 4096 ADC counts per \SI{2}{\volt}.}
    \label{fig:typ_wvfs}
\end{minipage}
\end{figure}

The left plot of Fig.~\ref{fig:typ_wvfs} shows example PMT pulses for $\SI{1}{pe}$, $\SI{2}{pe}$ and $\SI{3}{pe}$ at $\SI{-22}{\celsius}$. The SNR was determined to be 60 (62) for $\SI{-22}{\celsius}$ ($\SI{-38}{\celsius}$), clearly exceeding the required SNR of at least 25. The AFE response must also be linear in amplitude to the incident light intensity within $10\%$ in the dynamic range from $\SI{0.2}{pe}$ to $\SI{50}{pe}$. Example PMT pulses for large charges can be seen in the right plot of Fig.~\ref{fig:typ_wvfs} for $\SI{-22}{\celsius}$. The ADC dynamic range was found to extend from $\sim \SI{0.1}{pe}$ to $\sim \SI{70}{pe}$, saturating for charges beyond independent of the temperature. At $\SI{-22}{\celsius}$ linearity on the $10\%$ level is fulfilled throughout the dynamic range. Long-term (25 days) measurements of the discriminator threshold have also shown to remain stable, with a less than 0.05 mean pe drift per week.

\begin{figure}[tb]
\begin{minipage}{0.6\textwidth}
    \centering  
         \includegraphics{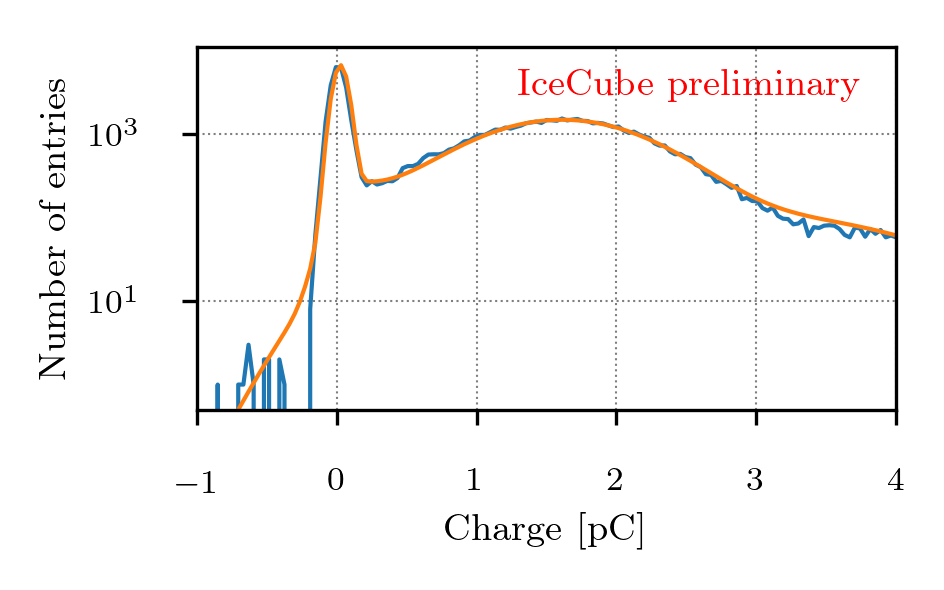}
\end{minipage}
\hfill
\begin{minipage}{0.4\textwidth}
    \caption{Single photo-electron charge histogram of an mDOM PMT as extracted by the mainboard MCU at $\SI{-35}{\celsius}$. A multi-component fit (\textit{orange}) allows to deduce the mean number of photoelectrons per pulse as well as the gain of the PMT.}
    \label{fig:charge_spectrum}
\end{minipage}
\end{figure}

The performance of the mDOM mainboard has been tested in a standalone setup, as well as in operation in a fully integrated mDOM (see Fig.~\ref{fig:mDOM}). 
A basic functional check-out test of the mDOM mainboard consists of a check of the DDR3 memory, the uploading, flashing and read-back of the FPGA firmware, and the various sensors (pressure, temperature, magnetometer, etc). 
In the case of an integrated mDOM, communication to the PMT bases is checked, alongside requirements such as the read-out of the per-PMT waveforms, PMT signal rates, or the distribution of the time between consecutive PMT signals. Finally, the calibration devices (see Sec.~\ref{sec:calibration_devs}) are tested. The MCU software allows for fast ADC and discriminator scans, efficient block read-out of waveforms, and a charge stamp readout from in-board waveform integration. An example of such a charge distribution is shown in Fig.~\ref{fig:charge_spectrum}. Waveforms were acquired using the discriminator trigger at a \SI{0.2}{pe} threshold, while the PMT was illuminated with SPE (single pe) level photons\footnote{On average $\ll 1$ photons per pulse reach the photocathode.} by the closest LED flasher in the mDOM.

\subsection{Calibration Devices}
\label{sec:calibration_devs}
For detector calibration, each mDOM is equipped with three cameras and four bright illumination LEDs (for details see \cite{icrc2019:cameras}), as well as ten flasher LEDs. The calibration devices are mounted on the inside of the internal support structure (see Sec.~\ref{sec:mechanics}) behind glass windows, which separate them from the optical gel and ensure an optical transition between the device and air as necessary for correct operation. For the flasher LEDs and illumination LEDs, latching mechanisms are printed into the support structure, while the heavier cameras are attached to their printed support posts with sheet metal screws. The flasher LEDs of type Roithner XRL-400-5O emit light at a wavelength of \SI[separate-uncertainty=true]{405(10)}{nm} at a maximum brightness of $10^9$ photons per pulse with a pulse width (FWHM) of \SI{7.5}{ns} (shorter at lower intensities). Each LED is mounted on its own PCB and is driven by a Kapustinsky driver. The five flashers in an mDOM hemisphere (four located between the equatorial and polar PMTs, one in the pole region; see Fig.~\ref{fig:mDOM} right) are linked in a daisy chain and can be addressed individually\footnote{Failures on a PCB, especially the connectors, could result in loss of all subsequent flashers in the daisy chain.}.

\subsection{Mechanical Structure}
\label{sec:mechanics}
The mDOM features a borosilicate glass pressure housing designed and tested to withstand pressures up to $\SI{700}{bar}$. It provides protection against the constant hydrostatic pressure in the deep-ice environment of $\sim \SI{250}{bar}$ as well as short-term pressure spikes during freeze-in after module deployment. 

A silicone-based two-component gel provides optical contact between the PMTs and the pressure vessel as well as cushioning against vibrations for the internal components. The chosen gel retains elasticity and optical transparency for temperatures ranging from room temperature down to $\SI{-45}{\celsius}$, the latter corresponding to surface temperatures at South Pole during deployment. Components in direct contact with the gel were tested for gel-curing compatibility. In the mDOM integration process the gel for each mDOM hemisphere is poured separately. Before potting, the gel components are thoroughly degassed at a few mbar. To minimize air entrapment during mixing and ensure a uniform mixing ratio and gel flow, a Meter Mix PAR 30 gear pump dispenser is used. Before curing, the gel is evacuated again to remove air inclusions in recesses or rough surfaces.

Inside the pressure vessel, the mDOM features an internal support structure to mount PMTs, calibration devices and readout-electronics. Its geometry features expansion joints to compensate for thermal shrinkage. 

\section{Status and Outlook}
\label{sec:status_outlook}
At the time of writing, the mDOM design is undergoing a Design Verification Test (DVT) campaign. In this procedure, ten fully functional mDOMs are being produced and, after an initial functional check-out (at room temperature as well as at $\SI{-40}{\celsius}$), are being submitted to a series of tests followed by a final check-out to verify the continuous functionality of the module. The test program recreates the impact of critical situations that can potentially lead to module failure. It features a shock test (tipping over of module inside a transportation box), a vibration test (land and air transport), a harness load test (maximum load during the lowering of a string of modules into the borehole), a thermal shock test (submerging a cold module into the warm borehole water) as well as a high pressure cycle (pressure spikes during freeze-in). Each of the tests is run by at least one of the ten DVT modules, with some modules undergoing multiple tests. In addition, the functionality of all mDOM components is assessed at temperatures as low as $\SI{-40}{\celsius}$ over a short period of time for two modules and over a long period of time for three modules. 

The DVT will also provide input for the optimization of the final module production process. Mass production of the mDOMs for IceCube Upgrade will be split between three sites. The support structures will be equipped with calibration devices in Münster. The final assembly will take place at DESY/Zeuthen and at Michigan State University. Deployment of the modules at the South Pole is foreseen for the Antarctic summer season of 2022/23.

\setlength{\bibsep}{2pt}
\bibliographystyle{ICRC}
\bibliography{references}

\providecommand{\href}[2]{#2}\begingroup\raggedright\begin{thebibliography}{10}

\bibitem{Aartsen2016:icecube}
{\bfseries IceCube} Collaboration, M.~G. Aartsen {\em et~al.}
  \href{http://dx.doi.org/10.1088/1748-0221/12/03/P03012}{{\em JINST}
  {\bfseries 12} no.~03, (2017) P03012}.

\bibitem{Abbasi2012:deepcore}
{\bfseries IceCube} Collaboration, R.~Abbasi {\em et~al.}
  \href{http://dx.doi.org/10.1016/j.astropartphys.2012.01.004}{{\em APP}
  {\bfseries 35} (2012) 615}.

\bibitem{PhysRevD.99.032007}
{\bfseries IceCube} Collaboration, M.~G. Aartsen {\em et~al.}
  \href{http://dx.doi.org/10.1103/PhysRevD.99.032007}{{\em Phys. Rev. D}
  {\bfseries 99} (Feb, 2019) 032007}.

\bibitem{PhysRevLett.120.071801}
{\bfseries IceCube} Collaboration, M.~G. Aartsen {\em et~al.}
  \href{http://dx.doi.org/10.1103/PhysRevLett.120.071801}{{\em Phys. Rev.
  Lett.} {\bfseries 120} (Feb, 2018) 071801}.

\bibitem{Ishihara2019:IC_Upgrade}
{\bfseries IceCube} Collaboration, A.~Ishihara
  \href{http://dx.doi.org/10.22323/1.358.1031}{{\em PoS} {\bfseries ICRC2019}
  (2021) 1031}.

\bibitem{Loehner2013:km3net_module}
{\bfseries KM3NeT} Collaboration, H.~L{\"o}hner {\em et~al.}
  \href{http://dx.doi.org/10.1016/j.nima.2012.11.049}{{\em Nucl.\ Inst.\ Meth.}
  {\bfseries A718} (2013) 513--515}.

\bibitem{Unland2019:pmts}
M.~A. Unland~Elorrieta {\em et~al.}
  \href{http://dx.doi.org/10.1088/1748-0221/14/03/P03015}{{\em JINST}
  {\bfseries 14} no.~03, (2019) P03015}.

\bibitem{icrc2021:pmt_testing}
{\bfseries IceCube} Collaboration, L.~Halve {\em PoS} {\bfseries ICRC2021}
  (these proceedings) 1056.

\bibitem{Crisler2010:COUPPBase}
M.~Crisler {\em et~al.},
  \href{http://dx.doi.org/10.1109/NSSMIC.2010.5873871}{``{The Chicagoland
  Observatory Underground for Particle Physics cosmic ray veto system},''} in
  {\em IEEE Nuclear Science Symposuim Medical Imaging Conference},
  pp.~808--812.
\newblock 2010.

\bibitem{icrc2019:cameras}
{\bfseries IceCube} Collaboration, W.~Kang {\em PoS} {\bfseries ICRC2021}
  (these proceedings) 1064.

\end{thebibliography}\endgroup

\clearpage
\section*{Full Author List: IceCube Collaboration}


\scriptsize
\noindent
R. Abbasi$^{17}$,
M. Ackermann$^{59}$,
J. Adams$^{18}$,
J. A. Aguilar$^{12}$,
M. Ahlers$^{22}$,
M. Ahrens$^{50}$,
C. Alispach$^{28}$,
A. A. Alves Jr.$^{31}$,
N. M. Amin$^{42}$,
R. An$^{14}$
% ...
\\

\noindent
% ...
$^{12}$ Universit{\'e} Libre de Bruxelles, Science Faculty CP230, B-1050 Brussels, Belgium \\
% ...
$^{14}$ Department of Physics and Laboratory for Particle Physics and Cosmology, Harvard University, Cambridge, MA 02138, USA \\
% ...
$^{17}$ Department of Physics, Loyola University Chicago, Chicago, IL 60660, USA \\
$^{18}$ Dept. of Physics and Astronomy, University of Canterbury, Private Bag 4800, Christchurch, New Zealand \\
% ...
$^{22}$ Niels Bohr Institute, University of Copenhagen, DK-2100 Copenhagen, Denmark \\
% ...
$^{28}$ D{\'e}partement de physique nucl{\'e}aire et corpusculaire, Universit{\'e} de Gen{\`e}ve, CH-1211 Gen{\`e}ve, Switzerland \\
% ...
$^{31}$ Karlsruhe Institute of Technology, Institute for Astroparticle Physics, D-76021 Karlsruhe, Germany  \\
% ...
$^{42}$ Bartol Research Institute and Dept. of Physics and Astronomy, University of Delaware, Newark, DE 19716, USA \\
% ...
$^{50}$ Oskar Klein Centre and Dept. of Physics, Stockholm University, SE-10691 Stockholm, Sweden \\
% ...
$^{59}$ DESY, D-15738 Zeuthen, Germany \\
% ...

\subsection*{Acknowledgements}

\noindent
USA {\textendash} U.S. National Science Foundation-Office of Polar Programs,
U.S. National Science Foundation-Physics Division,
U.S. National Science Foundation-EPSCoR,
Wisconsin Alumni Research Foundation,
Center for High Throughput Computing (CHTC) at the University of Wisconsin{\textendash}Madison,
Open Science Grid (OSG),
Extreme Science and Engineering Discovery Environment (XSEDE),
Frontera computing project at the Texas Advanced Computing Center,
U.S. Department of Energy-National Energy Research Scientific Computing Center,
Particle astrophysics research computing center at the University of Maryland,
Institute for Cyber-Enabled Research at Michigan State University,
and Astroparticle physics computational facility at Marquette University;
Belgium {\textendash} Funds for Scientific Research (FRS-FNRS and FWO),
FWO Odysseus and Big Science programmes,
and Belgian Federal Science Policy Office (Belspo);
Germany {\textendash} Bundesministerium f{\"u}r Bildung und Forschung (BMBF),
Deutsche Forschungsgemeinschaft (DFG),
Helmholtz Alliance for Astroparticle Physics (HAP),
Initiative and Networking Fund of the Helmholtz Association,
Deutsches Elektronen Synchrotron (DESY),
and High Performance Computing cluster of the RWTH Aachen;
Sweden {\textendash} Swedish Research Council,
Swedish Polar Research Secretariat,
Swedish National Infrastructure for Computing (SNIC),
and Knut and Alice Wallenberg Foundation;
Australia {\textendash} Australian Research Council;
Canada {\textendash} Natural Sciences and Engineering Research Council of Canada,
Calcul Qu{\'e}bec, Compute Ontario, Canada Foundation for Innovation, WestGrid, and Compute Canada;
Denmark {\textendash} Villum Fonden and Carlsberg Foundation;
New Zealand {\textendash} Marsden Fund;
Japan {\textendash} Japan Society for Promotion of Science (JSPS)
and Institute for Global Prominent Research (IGPR) of Chiba University;
Korea {\textendash} National Research Foundation of Korea (NRF);
Switzerland {\textendash} Swiss National Science Foundation (SNSF);
United Kingdom {\textendash} Department of Physics, University of Oxford.
\end{document}